\documentstyle[11pt,aaspp4,tighten]{article}

\newcommand{\grad}{\vec\nabla}

\newcommand{\vv}{\vec v}
\newcommand{\vk}{{\vec k}}

\newcommand{\vx}{\vec x}

\newcommand{\Vt}{\vec V_{\rm t}}

\newcommand{\vO}{\vec\Omega}

\newcommand{\p}{\partial}
\newcommand{\dt}{\partial_t}

\newcommand{\bA}{{\bf A}}
\newcommand{\bB}{{\bf B}}
\newcommand{\bC}{{\bf C}}
\newcommand{\bP}{{\bf P}}
\newcommand{\bI}{{\bf I}}
\newcommand{\half}{\frac{1}{2}}
\newcommand{\disp}{\dot{\cal E}}
\newcommand{\tav}{\overline}
\newcommand{\dy}{~\mbox{d}}
\newcommand{\be}{\begin{equation}}
\newcommand{\ee}{\end{equation}}

\begin{document}

\title{\sc Fast Tides in Slow Stars: The Efficiency of Eddy Viscosity}

\author{Jeremy Goodman \\
	Siang Peng Oh
}

\affil{Princeton University Observatory, 
	Princeton, NJ~08544}

\authoraddr{Jeremy Goodman
Princeton University Observatory, Peyton Hall, Princeton, NJ 08544.
e-mail: jeremy@astro.princeton.edu}
       
\slugcomment{Submitted to the Astrophysical Journal}

\begin{abstract}

Turbulent viscosity is believed to circularize and synchronize
binary orbits and to damp stellar oscillations.  It is also believed
that when the tidal period is shorter than the turnover time of the
largest eddies, turbulent viscosity is partially suppressed.  The
degree of suppression, however, is disputed.
We re-examine both of these beliefs via (i) direct perturbative 
calculations, linearizing the fluid equations on a turbulent background; 
and (ii) numerical integration of a chaotic dynamical system subject to 
periodic forcing.
We find that dissipation of rapid tides is severely suppressed.
Furthermore, circularization of late-type binaries does not occur
by turbulent convection if it occurs on the main sequence.

\medskip\noindent
{\it POPe-703.  Submitted to the Astrophysical Journal 12/19/96}

\end{abstract}



\section{Introduction}

Eddy viscosity in convection zones is presumed to dissipate
the shear associated with variable tides, and hence gradually to
circularize and synchronize close binaries [\cite{Z:66}, \cite{Z:77},
\cite{Z:92} and references therein].
This theory of tidal circularization has been
tested against binaries containing giant stars, and
the agreement is satisfactory (\cite{VP:96}).

This paper will focus on a theoretical controversy concerning
eddy viscosity whose outcome may have consequences not only for
tidal circularization, but also for stellar pulsations and
oscillations.
Assuming a Kolmogorov cascade of eddies, we have on scales
$\lambda$ smaller than the mixing length ($l$ that
\begin{eqnarray}
v_\lambda &\approx& \left(\frac{\lambda}{l}\right)^{1/3} v_l,
\nonumber\\
\tau_\lambda &\approx& \left(\frac{\lambda}{l}\right)^{-2/3} \tau_l,
\nonumber\\
\nu_\lambda &\approx& \left(\frac{\lambda}{l}\right)^{2/3} \nu_l,
\label{eq:Kolmogorov}
\end{eqnarray}
where $v_l$, $\tau_l$, and $\nu_l$ are the convective
velocity, turnover time, and effective viscosity on the
scale of the mixing length, which is comparabale to the
pressure scale height, $H$;
while $v_\lambda$, $\tau_\lambda$, and $\nu_\lambda$ are the
corresponding quantities on smaller scales $\lambda < l$.
If the  eddies on scales are space-filling, then the total 
turbulent viscosity $\nu_T\approx\nu_l$.
But when the period ($P$) of the tide or oscillation is less
than $\tau_l$, momentum transport by the large eddies is inhibited.
During half a period (or perhaps one should consider $P/2\pi$),
fluid in such an eddy travels no farther than $(P/2\tau_l)l$.
Zahn (1966, 1989) and \cite{ZB:89} therefore suppose that the eddy
viscosity is reduced by the same factor; that is, they take
\begin{equation}
\nu_\lambda = \frac{1}{3}\lambda v_\lambda
\min\left(P/2\tau_\lambda,1\right).
\label{eq:Zlim}
\end{equation}
Under this hypothesis, the largest eddies continue to dominate
the total viscosity,
\begin{equation}
\nu_{\rm T}=  \frac{1}{3}
v_l l \min\left[\left(\frac{P}{2\tau_l}\right),1\right].
\label{eq:Zred}
\end{equation}
\cite{GN:77} and \cite{GK:77} have taken a different view.
These authors argue
that the viscosity $\nu_\lambda$ on any given scale should 
be severely suppressed---perhaps by an exponential factor---when
$P\ll 2\pi\tau_\lambda$, so that
$\nu_{\rm T}$ is dominated by the largest eddies
whose turnover time is less than $P/2\pi$.
Since $\nu_\lambda\propto\tau_\lambda^2$ [cf. eq.~(\ref{eq:Kolmogorov}],
it follows that
\begin{equation}
\nu_{\rm T}= \frac{1}{3}
v_l l \min\left[\left(\frac{P}{2\pi\tau_l}\right)^2,~1\right].
\label{eq:Gred}
\end{equation}
GN's brief argument can be quoted in its entirety:
\begin{quote}
To appreciate this, it is necessary to recognize that turbulent
eddies have lifetimes which are comparable to their turnover times.
Thus, even though the large convective eddies move across distances
of order $l\tau_{\rm T}/\tau_{\rm c}$ in a tidal period, they do not
exchange momentum with the mean flow on this timescale.
The contribution to $\nu_{\rm T}$ made by the largest eddies is likely to
be smaller than $\nu_{\rm T~ max}$ by at least one additional factor of
$\tau_{\rm T}/\tau_{\rm c}$.
\end{quote}
Underlying this intuition, perhaps, is an analogy
with integrable hamiltonian systems, in which secular absorption 
occurs only through resonances; if such a system is perturbed rapidly compared
to any of its autonomous frequencies, no resonance with the perturbing
forces is possible.
Convective turbulence is neither integrable nor hamiltonian.
\cite{GM:91} have motivated GN's reduction factor by means
of an analogy with the kinetic theory of gases, in which they equate
$\tau_l$ to an intermolecular collision time.
Unfortunately, as shown in \S 4, 
\cite{VP:96}'s calibration against giants cannot be used to
decide between Eqs.~(\ref{eq:Zred}) and (\ref{eq:Gred}).

Another argument in favor of GN's prescription can be made by considering
the time inverse of absorption.
Consider a convecting star in an initially exactly
circular, short-period orbit.
Convection causes small fluctuations in the density distribution within
the star and corresponding departures of the exterior 
potential from symmetry; the latter modulate the binary orbit,
albeit at a very low level (\cite{P:92}).
If $\tau_l\ll P$, however, the energy of the orbit is adiabatically
invariant, so that no secular transfer of energy from the eddies
to the orbit can occur.
It would therefore seem to follow that the large eddies should not
absorb energy from a slightly eccentric short-period orbit.
Arguments from time-reversibility are very powerful
for thermal systems, but turbulent convection is not
reversible because of the flow of energy from large to small scales.
So this argument is not conclusive.

What is at stake?

\begin{description}
\item[Tidal circularization.]
In solar-type stars, $\tau_l\sim 20\dy$ in the middle
of the convection zone.
Thus $2\pi\tau_l$ is substantially longer than the orbital period
of circularized main-sequence binaries.
Furthermore, the turnover time increases with the depth of the convection zone
both on the main sequence towards lower masses, and in the pre-main-sequence
phase towards larger radii.
\cite{ZB:89}
therefore conclude that pre-main-sequence circularization would be
ineffective if the more severe reduction factor (\ref{eq:Gred}) were
correct.

\item[Stellar pulsation]
The fundamental pulsation period in variable stars
is $\sim(R^3/GM)^{1/2}$, which is usually shorter than $\tau_l$
unless the convection zone is thin. \cite{Gon:82} has argued
that the red edge of the instability strip is more consistent
with Zahn's prescription (\ref{eq:Zred}) than with GN's (\ref{eq:Gred}).

\item[Solar oscillations]
Goldreich and collaborators have constructed a theory for the
excitation and damping of the solar p-modes by convection
[\cite{GK:77}; \cite{GK:88}; \cite{GKM:94} and references therein].
This theory is generally in good agreement with the observed energies
of the modes, certainly more so than any alternative,
but it would have to be severely modified if eq.~(\ref{eq:Zred})
were applicable.
Acoustic emission by turbulence can be estimated, at least
roughly, with some confidence; if the damping processes were enhanced
then the energy in the convection would be lower than observed.

\end{description}

The outline of our paper is as follows.
In the next section we attack the fluid equations directly,
deriving linearized equations for the perturbation in the
turbulent velocities due to the tide.
The in-phase correlation of the perturbed velocity with the tidal force 
determines the rate of tidal work done on the turbulence.
In \S 3 we study a toy model for tidally perturbed convection:
a system of strongly coupled nonlinear oscillators driven by
a periodic external force.
By numerical integrations, we find the rate of absorption of energy
of this system as a function of the frequency of the external forcing.
In \S 4 we discuss the implications of our results for circularization
of late-type spectroscopic binaries.

\section{Perturbative Methods for Eddy Viscosity}

We divide the total velocity field into a large-scale
flow $\Vt$ associated with the equilibrium tide (or stellar
pulsation) and
a small-scale flow $\vv_c$ associated with the convection:
$\vv=\Vt+\vv_{\rm c}$.
The small-scale motions are approximately incompressible
and isentropic, so we describe them by the eulerian equations
\begin{eqnarray}
\dt\vv_c + \Vt\cdot\grad\vv_c+\vv_c\cdot\grad\Vt+\vv_c\cdot\grad\vv_c
+\grad w_c  &=& 0,\\
\label{eq:euler0}
\grad\cdot\vv_c &=& 0.
\label{eq:divv}
\end{eqnarray}
The quantity $w_c$ incorporates pressure and gravitational
accelerations, which are gradients of scalars in the
present approximation.
Buoyancy and thermal diffusion are not represented here,
although they are necessary to sustain convection; nor do
these equations incorporate coriolis forces or molecular viscosity.
We have considered these complications at length.
The conservative effects (buoyancy and coriolis forces)
do not seem to affect our conclusions at all,
although they complicate the algebra, and
the true dissipative terms are important only if they are
large enough to damp the tide directly.

We adopt a local approximation in which the lengthscales of the
convection are infinitesimal compared to those of the tide.
Locality is inherent in the notion of an effective viscosity and is 
responsible for much of the usefulness of the concept.
In this approximation, the tide couples to the turbulence
through the first-order spatial derivatives of $\Vt$.
Equation (\ref{eq:euler0}) contains the term
$\Vt\cdot\grad\vv_c$, however, in which $\Vt$ is not differentiated.
This term includes bulk transport of the local eddies, which
does not directly contribute to dissipation.
We eliminate this term by adopting a semi-lagrangian
coordinate mesh that deforms to follow the large-scale displacement.
Thus if $\{x^i|i\in 1,2,3\}$ are fixed coordinates, then
the semilagrangian system $\{x'^j\}$ is defined implicitly by
\begin{equation}
\left(\frac{\p x^i}{\p t}\right)_{\vx'} = V_{\rm t}^i (\vx,t).
\label{eq:defxp}
\end{equation}
Since we expect that only first spatial derivatives
of $\Vt$ are important for dissipation, we represent $\Vt$ as a linear
function of the coordinates,
\begin{equation}
\Vt=\bA(t)\cdot\vx ~~~\mbox{or}~~~V_{\rm t}^i = A^i_j(t)x^j,
\label{eq:linVt}
\end{equation}
in which the matrix $\bA$ depends upon time but not space.
The symmetric and antisymmetric parts of $\bA$ describe the local 
instantaneous shear and vorticity of the tidal flow.

Equations (\ref{eq:defxp})-(\ref{eq:linVt}) can be formally integrated
to obtain a time-dependent linear transformation between the two
coordinate systems:
\begin{equation}
\vx = \bB(t)\cdot\vx'~~~\mbox{or}~~~x^i = B^i_j x'^j.
\label{eq:trans}
\end{equation}
The matrix $\bB(t)$ is related to $\bA(t)$ by the differential
equation
\begin{equation}
\dot\bB(t) = \bA(t)\cdot\bB(t),
\label{eq:Bdot}
\end{equation}
which has the power-series solution
\begin{equation}
\bB(t) = 1+\int\limits_{-\infty}^t dt_1 \bA(t_1) 
+\int\limits_{-\infty}^t dt_1\int\limits_{-\infty}^{t_1}dt_2 
\bA(t_1)\cdot\bA(t_2)  +\ldots
\end{equation}
if $\vx=\vx'$ in the far past.
The contravariant components of the small-scale velocity field
in the semilagrangian system are related to the original
components by
\begin{equation}
v'^i \equiv \frac{\p x'^i}{\p x^j} v^j~~~\mbox{or}~~~
\vv' = \bB^{-1}\cdot\vv,
\label{eq:vtrans}
\end{equation}
and the equations of motion become
\begin{eqnarray}
\label{eq:eulerp}
\dt\vv' + 2\bB^{-1}\cdot\bA\cdot\bB\cdot\vv' +
\vv'\cdot\grad'\vv' +\grad'w &=&0,\\
\grad'\cdot\vv' &=& 0.
\label{eq:divp}
\end{eqnarray}
(The subscript ``c'' will be omitted from
the small-scale quantities henceforth.)
As a check, we note that if $\bA$ were purely antisymmetric,
then  $\bA\cdot\vec u=\vec\Omega\times\vec u$ for an any
vector $\vec u$, with $\vO(t)$ the instantaneous 
angular velocity; $\bB(t)=(\bB^T)^{-1}$ would be a rotation matrix;
and eq.~(\ref{eq:eulerp}) would reduce to the familiar form
\begin{displaymath}
\dt\vv' +2\vO'\times\vv' +\vv'\cdot\grad'\vv' +\grad' w = 0,
\end{displaymath}
with $\vO'\equiv\bB^{-1}\cdot\vO$.

The average rate of work done (per unit mass) 
on the small-scale velocities by the tide is
\begin{eqnarray}
\disp &=&\half\left\langle\frac{d}{dt}\tav{\vv\cdot\vv}\right\rangle=
\half\left\langle\frac{d}{dt}\tav{\vv'\cdot\bB^T\cdot\bB\cdot\vv'}\right\rangle
\nonumber\\
&=&\langle\tav{\vv'\cdot\bB^T\cdot\bB\cdot\dt\vv'}\rangle
+\half\langle\tav{\vv'\cdot\bB^T\cdot(\bA^T+\bA)\bB\cdot\vv'}\rangle
\label{eq:disp0}
\end{eqnarray}
Actually we intend a triple average: the angle brackets denote
an average over space and over realizations of the turbulence,
while the overline denotes an average over time.
Using equation (\ref{eq:eulerp}) and spatial integration by parts, one has
\begin{eqnarray}
\disp &=& -\half\langle\tav{\vv'\cdot\bB^T\cdot(\bA^T+\bA)\bB\cdot\vv'}\rangle
\nonumber\\
&=& -\half\langle\tav{\vv\cdot(\bA^T+\bA)\vv}\rangle.
\label{eq:avdisp}
\end{eqnarray}
This could have been derived directly from equation
(\ref{eq:euler0}) without use of the primed coordinates,
but the semilagrangian system proves convenient in the next step.
Henceforth let $\bA$ be symmetric, $\bA=\bA^T$,
since there should not be any dissipation
associated with the antisymmetric (rotational) part of the
the tidal velocity gradient.

The velocity $\vv$ or $\vv'$ appearing in the dissipation formula
(\ref{eq:avdisp}) is the actual velocity field in the presence of
the tide, not the the unperturbed turbulence $\vv_0$ that would
exist without it.
When the tidal shear is small, it is appropriate to evaluate
$\delta\vv'\equiv\vv'-\vv_0$ by perturbation theory
in ascending powers of $\bA$.
We can then express the dissipation rate (\ref{eq:avdisp}) in terms
of the statistical properties of $\vv_0$, even
though it is beyond our power to evaluate $\vv_0$ directly.
We assume that the unperturbed turbulent velocities can be 
described as a stationary random process with correlation function 
\begin{equation}
C^{mn}(\Delta \vec x,\Delta t) = \langle v_0^m(\vec x + \Delta\vec x,
t+\Delta t)v_0^n(\vec x, t)\rangle.
\label{eq:Cmn}
\end{equation}
We do {\it not} assume that $\vv_0(\vx,t)$ has gaussian statistics.

The first-order contribution to $\disp$ is obtained by replacing
$\vv'$ with $\vv_0$ in equation (\ref{eq:avdisp}).
This contribution vanishes, as can be demonstrated with
a temporal integration by parts.
At second order,
\begin{equation}
\disp_2 = -2\langle\tav{\vv_0\cdot\bA\cdot\delta\vv'}\rangle
\label{eq:E2}
\end{equation}
To evaluate this, one must calculate $\delta\vv'$ to
first order in the tide.
Linearizing equation (\ref{eq:eulerp}), 
\begin{equation}
\dt\delta\vv'(\vx',t)= -2\bA(t)\cdot\vv_0(\vx',t)-\vv_0\cdot\grad'\delta
\vv' - \delta\vv'\cdot\grad'\vv_0 ~~- \mbox{(pressure term)}.
\label{eq:dv0}
\end{equation}
The pressure term has not been written out because its only role
is to enforce $\grad'\cdot\delta\vv'=0$, which can be imposed
directly in Fourier space.
We will express $\delta\vv'$, $\vv_0$, and $\bA$ in terms of their
Fourier transforms:
\begin{eqnarray}
\delta\vv'(\vx',t)&=& \int\frac{d\omega}{2\pi}\int\frac{d^3\vk}{(2\pi)^3}
e^{i\vk\cdot\vx'-i\omega t}\delta\vv'(\vk,\omega)\nonumber\\
\vv_0(\vx',t)&=& \int\frac{d\omega}{2\pi}\int\frac{d^3\vk}{(2\pi)^3}
e^{i\vk\cdot\vx'-i\omega t}\vv_0(\vk,\omega)\nonumber\\
\bA(t)&=& \mbox{Real}\left[\bA(\Omega)e^{-i\Omega t}\right]
=\frac{1}{2}\left[\bA(\Omega)e^{-i\Omega t}+\bA(-\Omega)e^{i\Omega t}
\right].
\label{eq:FTdef}
\end{eqnarray}
Since the quantities on the left are real, their transforms satisfy
$\delta\vv'(-\vk,-\omega)=\delta\vv'(\vk,\omega)^*$,
$\vv_0(-\vk,-\omega)=\vv_0(\vk,\omega)^*$, and
$\bA(-\Omega)=\bA(\Omega)^*$.
When necessary, $\omega$ and $\Omega$ are considered to have an
infinitesimal positive imaginary part so that $\delta\vv'(\vx',t)$
vanishes as $t\to-\infty$.
Equation (\ref{eq:dv0}) becomes
\begin{eqnarray}
&-&i\omega\delta\vv'(\vk,\omega)= -\bP_{\vk}\cdot\left\{
\bA(\Omega)\cdot\vv_0(\omega-\Omega,\vk) 
+\bA(-\Omega)\cdot\vv_0(\omega+\Omega,\vk)
+\phantom{\frac{d\vk'}{(2\pi)^3}}\right.\nonumber\\
&&\left. \int\frac{d\omega'}{2\pi}\int\frac{d\vk'}{(2\pi)^3}
\left[\vv_0(\vk',\omega')\delta\vv'(\vk-\vk',\omega-\omega')
+\delta\vv'(\vk',\omega')\vv_0(\vk-\vk',\omega-\omega')\right]\cdot(i\vk)
\right\},\nonumber\\
\label{eq:FTdveqn}
\end{eqnarray}
where
\begin{equation}
\bP_\vk \equiv \bI - k^{-2}\vk\vk
\label{eq:Pdef}
\end{equation}
is a projection operator that maintains incompressibility,
\begin{equation}
\vk\cdot\delta\vv'(\vk,\omega)=0.
\end{equation}

The terms on the second line in equation (\ref{eq:FTdveqn})
are small compared to those on the first line.
Neglecting the second line, one has $\delta v'\sim A V_c \Omega^{-1}$,
where $V_c$ is typical velocity for the convection.
It follows that the second line is smaller than the first by
a factor $\sim V_c/(\Omega L_c)= (\Omega\tau_c)^{-1}$, where 
$L_c$ is a typical convective lengthscale, and $\tau_c=L_c/V_c$ is
the corresponding eddy turnover time.
Note that we now have two characteristic
dimensionless parameters: the first is
the tidal strain, $\Omega^{-1}|\bA|$, which is always small
in detached binaries; the second, $(\Omega\tau_c)^{-1}$, is completely
independent of the first and may or may not be small.
When the latter is small,
we may formally solve equation (\ref{eq:FTdveqn}) as a power series in
$(\Omega\tau_c)^{-1}$.
Let $\delta_{1,n}\vv'$ denote the contribution to $\delta\vv'$ that
is first order in the tidal strain and $n^{\rm th}$ order in
$(\Omega\tau_c)^{-1}$.
Successive terms are generated recursively: substituting
$\sum_{k=1}^n\delta_{1,k}\vv'$ for $\delta\vv'$ in the second line of
(\ref{eq:FTdveqn}), one solves for $\sum_{k=1}^{n+1}\delta_{1,k}\vv'$.

To lowest order in $(\Omega\tau_c)^{-1}$, we have
\begin{equation}
\delta_{1,1}\vv'(\vk,\omega)= -\frac{i}{\omega}\bP_{\vk}\cdot\left[
\bA(\Omega)\cdot\vv_0(\omega-\Omega,\vk) 
+\bA(-\Omega)\cdot\vv_0(\omega+\Omega,\vk)\right],
\label{eq:lowest}
\end{equation}
and upon substituting this into equation (\ref{eq:E2}), we have
\begin{eqnarray}
\disp_{2,2} &=& \int\frac{d\omega}{2\pi}\int\frac{d^3\vk}{(2\pi)^3}
\left(\frac{i}{\omega}\right)\mbox{Trace}[\bC(\vk,\omega+\Omega)\cdot
\bA(\Omega)\cdot\bP_{\vk}\cdot\bA^*(\Omega)+\nonumber\\
&&\bC(\vk,\omega-\Omega)\cdot\bA^*(\Omega)\cdot\bP_{\vk}\cdot\bA(\Omega)].
\label{eq:form1}
\end{eqnarray}
We have made use of
\begin{equation}
\langle\vv_0(\vk',\omega')\vv_0^T(\vk,\omega)\rangle=
(2\pi)^4\delta^3(\vk'+\vk)\delta(\omega+\omega')\bC(\vk,\omega),
\label{eq:FTC}
\end{equation}
where $\bC(\vk,\omega)$ is the Fourier transform of the correlation
function (\ref{eq:Cmn}), and the transpose symbol is used for 
clarity: $\vv\vv^T$ is a matrix with components $v^i v^j$.
The delta functions in equation (\ref{eq:FTC}) result from
stationary statistics: that is, from the fact that the correlation
function (\ref{eq:Cmn}) depends only on the differences 
between the two points at which the turbulent
velocity is correlated but not on the absolute positions and times.

The subscripts follow the scheme described above:
that is, corresponding to the term $\delta_{1,n-1}$ in the expansion
of the perturbed velocity, there is through equation (\ref{eq:E2})
a contribution $\disp_{2,n}$ to the dissipation rate that is second
order in the tide and $n^{\rm th}$ order in the convection.
In addition to two powers of $\bA$ and $n$ powers of $\vv_0$,
$\disp_{2,n}$ also contains $(n-2)$ spatial derivatives applied to
$\vv_0$ and $(n-1)$ time integrations.
Assuming that each differentiated factor of $\vv_0$ is of order
$\tau_c^{-1}$, and that each time integration donates a factor $\Omega^{-1}$,
we may expect that
\begin{equation}
\disp_{2,n}\sim |\bA|^2 V_c^2\tau_c (\Omega\tau_c)^{1-n}.
\label{eq:naive}
\end{equation}
Since the first nonvanishing term is $\disp_{2,2}$, the above scaling
agrees with the prescription (\ref{eq:Zred}) of Zahn.
It will now be shown, however, that $\disp_{2,2}$ is actually much smaller
than predicted by the scaling (\ref{eq:naive}).

The trace in equation (\ref{eq:form1}) is real because
\begin{displaymath}
\mbox{Trace}(\bC\cdot\bA\cdot\bP_{\vk}\cdot\bA^*)=
\mbox{Trace}(\bC\cdot\bA\cdot\bP_{\vk}\cdot\bA^*)^T =
\mbox{Trace}(\bA^{*}\cdot\bP_{\vk}^*\cdot\bA\cdot\bC^*)=
\mbox{Trace}(\bC^*\cdot\bA^{*}\cdot\bP_{\vk}^*\cdot\bA),
\label{eq:form2}
\end{displaymath}
in which the arguments of the various matrices have been
temporarily suppressed, and we have made use of $\bA=\bA^T$,
$\bP_{\vk}^T=\bP_{\vk}^*$ and
$\bC(\vk,\omega)^T=\bC(\vk,\omega)^*$.
On the other hand, the term $i\omega^{-1}$ in eq. (\ref{eq:form1})
appears to be purely imaginary.
Since $\disp_2$ is real by definition [eqn. (\ref{eq:E2})],
it would seem that $\disp_{2,2}$ must vanish.
As noted above, however, all frequencies should be endowed with an
infinitesimal positive imaginary part, because one wants the particular
solution for which $\delta\vv'$ vanishes as $t\to-\infty$.
Hence one must interpret $i/\omega$ as
\begin{equation}
\lim_{\epsilon\to 0} \frac{i}{\omega+i\epsilon}
= \lim_{\epsilon\to 0} \frac{i\omega}{\omega^2+\epsilon^2}
~+~ \lim_{\epsilon\to 0} \frac{\epsilon}{\omega^2+\epsilon^2}.
\label{eq:limits}
\end{equation}
The first term on the right is indeed imaginary.
The second term is real and represents $\pi\delta(\omega)$.
This term gives a nonzero result for $\disp_{2,2}$, which can be written
using $\bA(-\Omega)=\bA^*(\Omega)$ as
\begin{equation}
\disp_{2,2} = \half\int\frac{d^3\vk}{(2\pi)^3}
\mbox{Trace}[\bC(\vk,+\Omega)\cdot
\bA(\Omega)\cdot\bP_{\vk}\cdot\bA^*(\Omega)+
\bC(\vk,-\Omega)\cdot\bA(-\Omega)\cdot\bP_{\vk}\cdot\bA^*(-\Omega)].
\label{eq:reson}
\end{equation}

We can compare the last result with the dissipation rate provided
by a true molecular viscosity $\nu$:
\begin{equation}
\disp_{\rm visc} = \frac{1}{2}\nu\mbox{Trace}[\bA(\Omega)\cdot\bA^*(\Omega)].
\label{eq:truevisc}
\end{equation}
In general, the effective viscosity provided by equation (\ref{eq:reson})
is anisotropic.
If the turbulence is isotropic---though there is no good reason that
convective turbulence should be---then
\begin{equation}
\bC(\vk,\omega)= C(k,\Omega)\bP_{\vk},
\label{eq:isotropic}
\end{equation}
with $C(k,-\Omega)=C(k,\Omega)$.
In this case, equation (\ref{eq:reson}) can be simplified to
\begin{equation}
\disp_{2,2} \to \frac{7}{15}\mbox{Trace}[\bA(\Omega)\cdot\bA^*(\Omega)]
\int\frac{d^3\vk}{(2\pi)^3} C(k,\Omega).
\label{eq:special}
\end{equation}
We have taken $\mbox{Trace}(\bA)=0$, since all of our derivations
assume incompressible motion; our arguments would need to be generalized
to treat damping of compressive motions, such as solar $p$ modes.
The integral in equation (\ref{eq:special}) is the kinetic energy
of the turbulence per unit frequency per unit mass
at the frequency of the tide.
Denoting this quantity by ${\cal E}(\Omega)$, we can express the turbulent
effective viscosity in the isotropic case as
\begin{equation}
\nu_T \sim \frac{14}{15}{\cal E}(\Omega),
\label{eq:nuT2}
\end{equation}
where ${\cal E}(\omega)$ is normalized so that the average kinetic
energy per unit mass is
\begin{equation}
\left\langle\frac{1}{2}|\vv(\vx,t)|^2\right\rangle=
\int\limits_{-\infty}^{\infty}\frac{d\omega}{2\pi}{\cal E}(\omega).
\label{eq:pspectrum}
\end{equation}
The ``$\sim$'' symbol acknowledges that
this is only the first term of an expansion in powers of 
$(\Omega\tau_c)^{-1}$.

Whether one considers the isotropic form (\ref{eq:special}) or the more
general form (\ref{eq:reson}), {\it the leading-order
contribution to the turbulent dissipation rate depends entirely upon
the power spectral density of the turbulent velocities at the frequency 
of the tide}.
Imagine an artificial turbulence in which the eddies on all scales
had correlation times $\tau_c\gg\Omega^{-1}$.
Then $\bC(\vk,\pm\Omega)$ in equation (\ref{eq:reson}) and
${\cal E}(\pm\Omega)$ in equation (\ref{eq:nuT2}) would be exponentially
small, and $\disp_{2,2}\propto\exp[-(\Omega\tau_c)]$.

In fact, even if the largest eddies are much slower than the tide, 
there will be smaller and less energetic eddies with
decorrelation times $\tau_\lambda\le\Omega^{-1}$.
Our perturbative expansion is invalid for these small eddies,
which correspond to large values of the wavenumber $\vk$ in 
eq.~(\ref{eq:reson}).
It seems likely, therefore, that the expansion in powers of
$(\Omega\tau_c)^{-1}$ is at best asymptotic rather than convergent.

There is a question about how to translate the heuristic Kolmogorov
model into a prediction for ${\cal E}(\omega)$.
In the usual interpretation, the turbulent power per unit frequency at
frequency $\omega$ is due to eddies of size $\lambda$ such that
their turnover time $\tau_\lambda\sim\omega^{-1}$.
The standard Kolmogorov scalings (\ref{eq:Kolmogorov}) applied
to eq. (\ref{eq:reson}) then lead to GN's prescription 
(\ref{eq:Gred}) up to a dimensionless constant of order unity.
Another interpretation is possible, however, and leads to a prescription
intermediate between equations (\ref{eq:Gred}) and (\ref{eq:Zred}).
The correlation function whose Fourier transform appears in the dissipation
rate (\ref{eq:reson}) is measured at fixed spatial positions.
A small eddy of size $\lambda\sim k^{-1}$ is carried past a fixed point
by the resultant velocity of all larger eddies in which it is embedded;
this velocity is $\sim v_l$, the velocity of the largest eddies.
The apparent frequency of the small eddy measured by a fixed observer 
is $k v_l$, which is larger than the turnover rate $\tau_\lambda^{-1}$
by a factor $\sim (kl)^{1/3}$ [cf. eq. (\ref{eq:Kolmogorov})].
According to this logic and our result (\ref{eq:nuT2}), the effective
turbulent velocity scales with the tidal period as
\begin{equation}
\nu_{\rm T}= v_l l \min\left[\left(\frac{P}{2\pi\tau_l}\right)^{5/3},~1\right].
\label{eq:hybrid}
\end{equation}
Physically, however, we do not see how the mere fact that
a small eddy is borne along by a larger one should affect its rate of
absorption of energy from a spatially uniform tidal shear.
Perhaps higher-order terms replace
the eulerian power spectrum $\bC(\vk,\Omega)$ by a lagrangian
equivalent that measures the decorrelation rate of the turbulence following
the fluid, in which case one reverts to the earlier interpretation,
and the exponent $5/3$ is replaced by $2$, in agreement with eq. 
(\ref{eq:Gred}).
Even if eq.~(\ref{eq:hybrid}) is correct, it is closer
to eq.~(\ref{eq:Gred}) than to Zahn's prescription (\ref{eq:Zred}).
In the rest of our paper, therefore, we adopt GN's prescription.

\section{Chaotic Dynamical Models}

As a toy model for convection, we have considered a chain of coupled
nonlinear oscillators, which may loosely be considered to be
interacting vortices. The lagrangian of the
unforced system is
\begin{equation}
L= \sum_{i=0}^{N-1} \, \frac{1}{2} \dot{y}_{i}^{2}
- \frac{1}{3} (y_{i+1}-y_{i})^{3} - \frac{1}{4} (y_{i+1}-y_{i})^{4}
\label{eq:chain}
\end{equation}
with fixed end-point boundary conditions $y_{0}=y_{N}=0$.
The linearized normal modes of this system have zero frequency,
so their interactions at finite amplitude should be analogous
to strong turbulence.

We have simulated this system (\ref{eq:chain}) by direct numerical
integration of the equations of motion for $N=63$.
We set the initial displacement of all oscillators to zero and
assign velocities drawn independently from a
gaussian distribution of unit variance,
after which the velocities are rescaled so that the total energy
\begin{equation}
E\equiv \sum_{i=0}^{N-1} \, \frac{1}{2} \dot{y}_{i}^{2}
+ \frac{1}{3} (y_{i+1}-y_{i})^{3} + \frac{1}{4} (y_{i+1}-y_{i})^{4}
\label{eq:enerdef}
\end{equation}
is $N-1$ at the start of all runs.

To verify that the system is chaotic, we estimate
the maximum Lyapunov exponent
(cf. \cite{AFH:94})
\begin{equation}
\Lambda\equiv\lim\limits_{t\to\infty}\log_2\left[\frac{d(t)}{d(0)}\right],
\end{equation}
where $d$ is the infinitesimal distance in phase space between
neighboring trajectories, and the logarithms to base 2 reflect the
convention that information regarding the initial state is measured in
bits.  We solve the non-linear equations of motion for a reference
trajectory $Y(t)\equiv\{y_1(t),\ldots,y_{N-1}(t)\}$,
\begin{equation}
\dot Y(t)=F[Y(t)],
\end{equation}
and the linearized variational equations describing the displacement
of a neighboring trajectory,
\begin{equation}
\dot{\delta Y }(t) = \left. \frac{\partial F}{\partial Y} \right|_{Y(t)}
\cdot\delta Y (t).
\end{equation}
The distance between neighboring trajectories is measured by
\begin{equation}
d(t)\equiv \Vert \delta Y (t)\Vert = \left|\sum\limits_{i=1}^{N-1}
\delta y_i(t)^2 \right|^{1/2}.
\end{equation}
Although the limit $t\to\infty$ is inaccessible, we find the
increase in $d(t)$ to be accurately exponential and estimate
$\Lambda\approx 2.7$.

\begin{figure}
\plotone{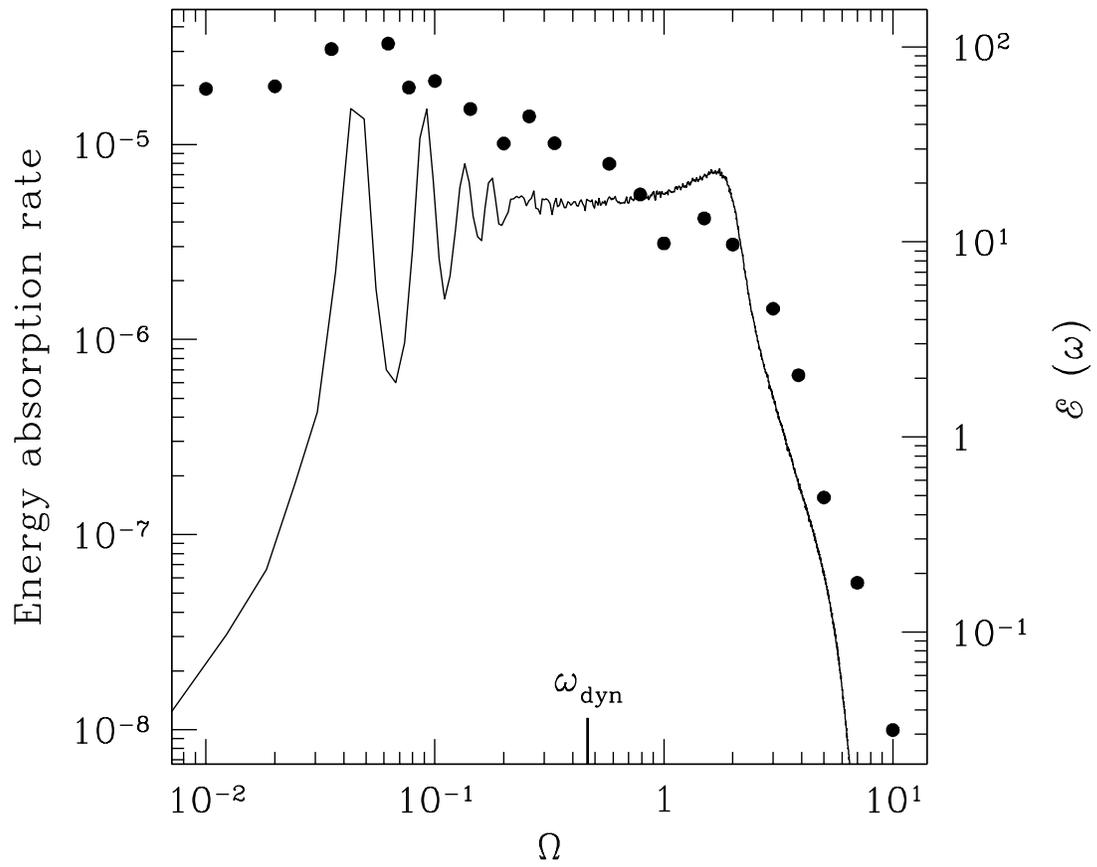}
\caption{Energy absorption rate vs. forcing frequency
(points), as compared with the velocity power spectrum
eq.~(\ref{eq:pspectrum}) (line).}
\label{cap:de}
\end{figure}

A quantity analogous to the energy per unit mass per unit frequency
of convection [eq.~(\ref{eq:pspectrum})],
is the temporal power spectrum of the oscillator velocities,
\begin{eqnarray}
{\cal E}(\omega)&=&
\label{eq:oscspec}
\frac{1}{N-1}\left\langle\left|\sum\limits_{i=1}^{N-1}\tilde v_i(\omega)
\right|^2\right\rangle,\nonumber\\
\tilde v_i(\omega)&\equiv&\int e^{i\omega t}\dot y_i(t),\nonumber
\end{eqnarray}
which we compute by discrete Fourier transforms and average over many
runs.  The spectrum (Fig.~\ref{cap:de}) has broad peaks at low
frequencies and a smoothly declining tail above a characteristic
frequency defined by
\begin{equation}
\omega_{\rm dyn}\equiv\left[\frac{\langle\Vert\dot Y\Vert^2\rangle}
{\langle\Vert Y\Vert^2\rangle}\right]^{1/2} \approx 0.465.
\end{equation}
Equipartition of energy appears to hold among
the spatial Fourier modes defined by
\begin{equation}
q_k(t)\equiv \left(\frac{2}{N-1}\right)^{1/2}\sum\limits_{i=1}^{N-1}
y_i(t)\sin\left(\frac{\pi k i}{N}\right).
\label{eq:Fmodes}
\end{equation}
Although the eigenfrequencies of these modes vanish,
their characteristic frequencies at finite amplitude,
$\omega_{k,\rm dyn}\equiv\langle\dot q_k^2\rangle^{1/2}/
\langle q_k^2\rangle^{1/2}$, are found to be approximately proportional
to wavenumber $k$, and and the largest of these frequencies
are comparable with $\omega_{\rm dyn}$.

To simulate the tide, we add the force term
\begin{equation}
F_{j} = \epsilon \dot{y_{j}}\cos(\Omega t)
\end{equation}
to the equation of motion of a single oscillator ($j=22$).
The combination $\epsilon\cos(\Omega t)$ is analogous
to the tidal shear $\bA(t)$ of \S 2, and we usually
take $\epsilon\sim 0.05\omega_{\rm dyn}$.

To determine the secular energy absorption rate $\langle\dot
E\rangle$, the total energy $E(t)$ defined by eq.~(\ref{eq:enerdef})
is recorded as a time series, smoothed with a low-pass filter, and fit
to a straight line.  The results for $\langle\dot E\rangle$ as a
function of $\Omega$ are compared in Figure~\ref{cap:de} with
the power spectrum (\ref{eq:oscspec}).  The energy absorbed from the
``tide'' drops off sharply when $\Omega>\omega_{\rm dyn}$, as
expected.

Since our oscillator system is hamiltonian, the argument from
time-reversability in \S1 is applicable to it.  
That the system can absorb very little energy from a 
high-frequency tide therefore should not come as a surprise.
In fact, the fluctuation-dissipation theorem is applicable here and
predicts that energy is absorbed at a rate propotional to ${\cal E}(\omega)$
(cf. \cite{LL:80}, \cite{NT:97}).
It would be interesting to allow for
dissipation at large $k$ and stochastic forcing at small $k$
[cf. eq.~(\ref{eq:Fmodes}) above] in the
unperturbed oscillator system; this would
break time-reversal symmetry, establish a cascade from large scales
to small, and bring the model closer to real convective turbulence.
But to achieve a good separation of timescales would require many more
than 64 oscillators and might be a challenging computational exercise.

\section{Application to Late-Type Binary Stars}

In this paper, we have revisited the question of how efficiently
an oscillatory large-scale shear is dissipated by turbulence
when the reduced period of the shear ($P/2\pi$)
is shorter than the correlation
time of the energy-bearing eddies ($\tau_c$).
By a direct perturbative approach proceeding from the dynamical
equations of incompressible turbulence (\S 2), we find that the
dissipation rate is reduced by a factor scaling as $(P/2\pi\tau_c)^2$
[or perhaps $(P/2\pi\tau_c)^{5/3}$, depending on what one takes for
the temporal correlation of the turbulence on timescales
$\tau < \tau_c$].
The quadratic scaling is also found in numerical simulations
of a forced nonlinear oscillator chain (\S 3).
Our results therefore agree substantially with the quadratic
suppression (\ref{eq:Gred}) rather than 
the linear one (\ref{eq:Zred}).

The distinction between these prescriptions is
important astrophysically because it is often the case that
$P\ll 2\pi\tau_c$.
In the remainder of this paper, we apply our results to the 
circularization of late-type binaries.
Our treatment follows that of Zahn (1989, henceforth Z89),
except that we adopt the quadratic inefficiency factor rather than
the linear one.

\subsection{The long-period limit}

To establish a baseline for later discussion,
we first recalculate the circularization times of
late-type main-sequence binaries as though their
periods were long compared to all eddy turnover times.
Even at this naive level, there appears to be a discrepancy
between the tidal theory and the observed maximum period at
which main-sequence binaries are circularized.
When we take into account the inefficiency of slow convection, in
the next subsection, the discrepancy will become severe.

Consider a binary consisting of two stars identical to the present sun.
The circularization rate follows from equation (21) of Z89:
\begin{equation}
t_{\rm circ}^{-1}\equiv -\frac{d}{dt}\ln e = 84
\frac{\lambda}{t_f}\left(\frac{R}{a}\right)^8,
\label{eq:crate0}
\end{equation}
where
\begin{equation}
t_f\equiv\left(\frac{MR^2}{L}\right)^{1/3}
\label{eq:tfric}
\end{equation}
is Zahn's ``friction time,'' and the dimensionless
parameter $\lambda$ is defined by
\begin{equation}
\lambda = 205.6\frac{t_f}{MR^8}\int\limits_{r_c}^R\rho\nu_T~ r^8 dr,
\label{eq:lamdef}
\end{equation}
where $r_c$ is the inner radius of the convection zone.
The friction time is only very weakly dependent on
stellar mass and spectral type; for the sun, $t_f=0.432~\mbox{yr}$.
The parameter $\lambda$ is a volume average of
the dynamic turbulent viscosity, $\rho\nu_T$, weighted by
the square of the tidal shear; the radial displacement
of the equilibrium tide varies approximately as $r^4$.
We have multiplied Zahn's expression
for the circularization rate by a factor of two to account for the
dissipation in both stellar envelopes.

Using mixing-length theory, Z89 derives
\begin{equation}
\rho\nu_T^{(0)}
\approx \frac{4}{75}\frac{M}{4\pi Rt_f}\left(\frac{6c}{5}\right)^{1/3}
\alpha^{4/3} E^{2/3} x^{-2/3}(1-x)^2.
\label{eq:dynvisc}
\end{equation}
Here $\alpha$ is the usual mixing-length parameter (the ratio $l/H$ of
mixing length to pressure scale height), and $x\equiv r/R$ is the
fractional radius.
The superscript on $\nu_T^{(0)}$ indicates that this estimate of the
turbulent viscosity assumes $\Omega\tau_c\ll 1$.
The dimensionless parameter $c\le 1$ specifies
how much of the potential energy released by convection is converted
to kinetic energy on the scale of the mixing length; after discussing
the choices made by other authors, Zahn takes
$c= 9/32$, but the dissipation rate is not very sensitive to
$c$ because of the cube root.
The constant $E$ describes the run of density with radius in a
polytropic approximation to the convection zone:
\begin{equation}
\rho(r)\approx E\frac{M}{4\pi R^3}\left\{\frac{1}{n+1}\frac{R}{GM}
\left[\Phi(R)-\Phi(r)\right]\right\}^n
\approx E\left[\frac{2}{5}\left(\frac{1-x}{x}\right)\right]^{3/2},
\end{equation}
where $n\approx 3/2$ is the polytropic index.
Comparing the final expression above with the tabulated model
of \cite{BP:95}, one finds that $E\approx 6.1$ at the base of
the convection zone, which occurs at $0.712 R_{\sun}$, and that
$E$ decreases slowly to $\approx 5.4$ at the outermost tabulated point
($0.9464 R_{\sun}$).

We therefore find for the circularization time
\begin{equation}
t_{\rm circ}\approx 9.0\times 10^{10}\left(\frac{\alpha}{2}\right)^{-4/3}
\left(\frac{E}{6}\right)^{-2/3}
\left(\frac{\bar\rho}{\bar\rho_{\sun}}\right)^{8/3}
\left(\frac{P}{10~\rm d}\right)^{16/3} ~\mbox{yr},
\label{eq:tcirc0}
\end{equation}
where $R/a$ has been eliminated in favor of the orbital period.
As this formula shows, $t_{\rm circ}$ is sensitive to the stars' 
mean density $\bar\rho\equiv 3M/4\pi R^3$.

For a stellar population of age $t$,
the transition between circular and eccentric
orbits should occur at a period $P_{\rm circ}$ such that
\begin{equation}
t_{\rm circ}(P_{\rm circ})\approx \frac{1}{3} t,
\end{equation}
so that mean initial eccentricities of order $0.5$
will have been reduced to a few percent or less.
By this criterion, 
\begin{equation}
P_{\rm circ}\approx 5.9 \left(\frac{\alpha}{2}\right)^{1/4}
\left(\frac{E}{6}\right)^{1/8}
\left(\frac{\bar\rho}{\bar\rho_{\sun}}\right)^{-1/2}
\left(\frac{t}{16~\mbox{Gyr}}\right)^{3/16}~\mbox{d}
\label{eq:Pcirc0}
\end{equation}

Recall that eq.~(\ref{eq:Pcirc0}) is only an upper bound because it
neglects the inefficiency of slow convection; yet observations
suggest that $P_{\rm circ}$ exceeds this bound.
The halo-binary sample
of \cite{L:92} indicates $P_{\rm circ}\gtrsim 12~\mbox{d}$
for metal-poor, high-proper-motion dwarfs
($\mu\gtrsim 0.\arcsec27~\mbox{yr}^{-1}$, $[\mbox{Fe/H}]<-1.6$).
\cite{L:92} conclude  $P_{\rm circ}\approx 19~\mbox{d}$ based
principally on a binary with period $18.74~\mbox{d}$ and eccentricity 
$0.043\pm 0.021$.
This system is a single-lined spectroscopic binary, so that its
companion might be evolved.
But the sample also contains two cicular double-lined systems
with $P=10.74$ and $11.73~\mbox{d}$.
Based on colors, metallicities, and proper motions, 
all four stars are clearly on the main sequence and well below the turnoff
(\cite{LCL:88}).
\footnote{From the most metal-poor 16-Gyr theoretical isochrone of
\cite{BV:92}, which fits M92, the mean densities
of the stars in these two binaries are approximately $3.6$ and
$4.8$ times larger than that of the sun if they lie on the
main sequence.  Eq.~(\ref{eq:tcirc0}) then predicts
$t_{\rm circ}\approx 2.7\times 10^{12}$ and $6\times 10^{12}~\mbox{yr}$, 
respectively.
If the stars were giants, then they would be
$5.4$ and $8.2$ magnitudes more luminous, would be much more distant,
and would have absurdly large transverse velocities.}
Also, $P_{\rm circ}\approx 10~\mbox{d}$ has been estimated
for main-sequence binaries in the old open cluster M67 
(\cite{MM:88}, \cite{MLG:90}).

\subsection{Inefficient turbulence}

Revising the theoretical circularization time to allow
for inefficient dissipation of rapidly-varying tides
makes the conflict between theory and observation still more
acute.

The rapidity of the tide relative to the convection is characterized
globally by the dimensionless quantity
\begin{equation}
\eta' \equiv \frac{2\pi t_f}{P_{\rm tide}}\approx 99
\left(\frac{t_f}{t_{f,\sun}}\right)\left(\frac{10~\mbox{d}}{P_{\rm tide}}
\right).
\label{eq:etadef}
\end{equation}
which we have defined by analogy with Z89's parameter 
$\eta\equiv 2 t_f/P_{\rm tide}=\eta'/\pi$.
Locally at each radius within the convection zone, however, the
convective timescale is
\begin{equation}
\tau_l(r)\equiv\frac{l}{v_l} =\frac{l^2}{3\nu_T^{(0)}},
\end{equation}
where $l=\alpha H$ is the mixing length and $\nu_T^{(0)}$ is the turbulent
viscosity in the long-period limit [eq.~(\ref{eq:dynvisc})].
Where $\tau_l(r)>P_{\rm tide}/(2\pi)$, we apply
GN's prescription (\ref{eq:Gred}), so that the dynamical viscosity becomes
\begin{equation}
\rho\nu_T^{(\eta')} = \eta'^{-2}\frac{M}{4\pi R t_f} c x^{-3}(1-x)^{-1}.
\label{eq:dynvisc1}
\end{equation}
Taking $c=9/32$ as before, we use the smaller of the expressions
(\ref{eq:dynvisc}) and (\ref{eq:dynvisc1})
the integral (\ref{eq:lamdef}) for $\lambda$.
Notice that the reduced viscosity (\ref{eq:dynvisc1}) is independent
of the mixing length and of the density parameter $E$.
Where eq.~(\ref{eq:dynvisc1}) applies, in the lower parts of the convection
zone, the total dissipation rate receives equal contributions
from equal logarithmic intervals in exterior mass, $M-M_r$
(cf. GN).

\begin{figure}
\plotone{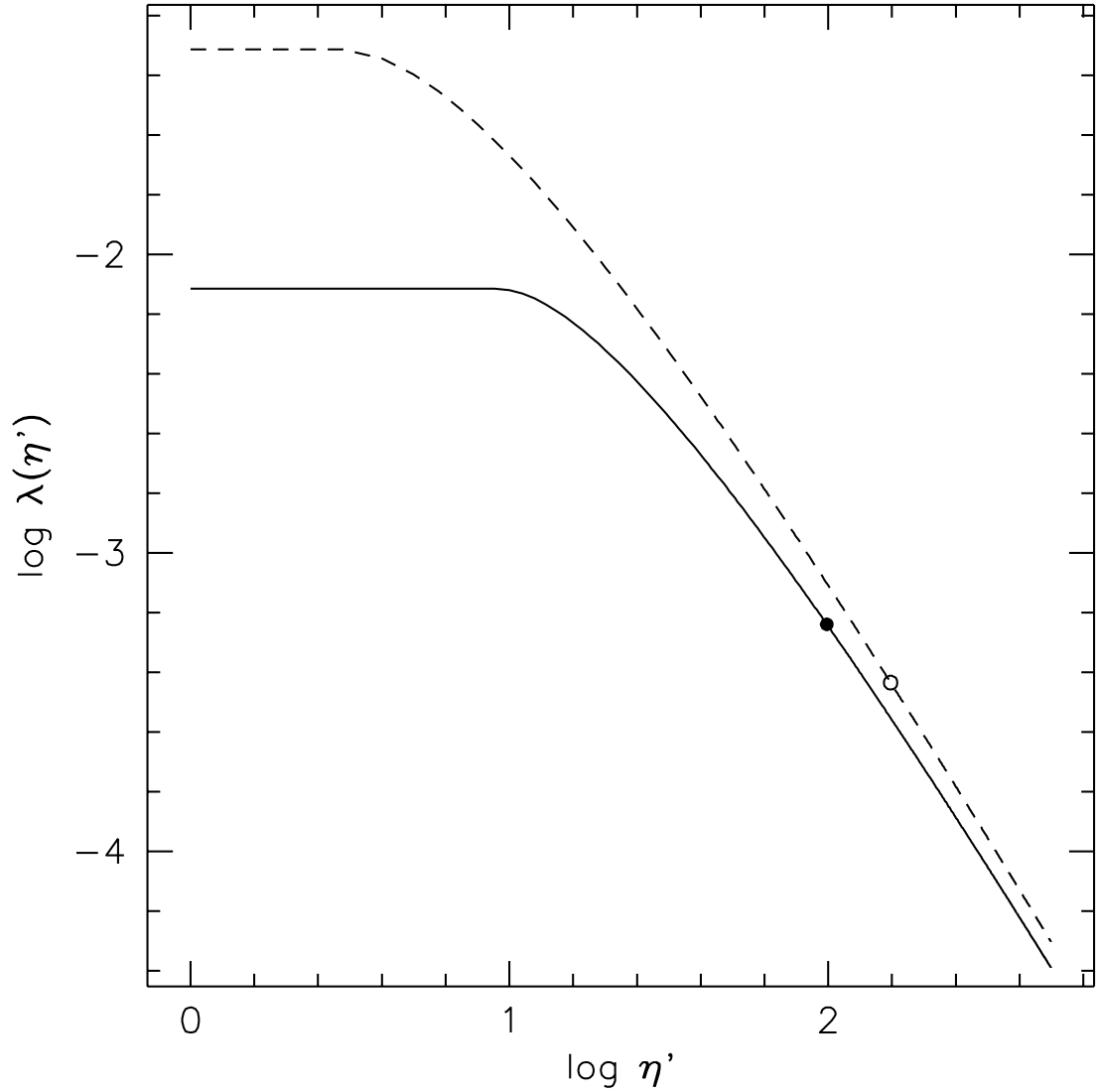}
\figcaption{Tidal dissipation parameter
$\lambda$ [eq.~(\ref{eq:lamdef})] versus dimensionless tidal frequency
$\eta'$ [eq.~(\ref{eq:etadef})].  Solid line: solar model.  Dashed
line: fully convective star.  Circles: two suns (solid) and two
protosuns (open) in a $10\dy$ orbit (see text).}
\label{cap:lam}
\end{figure}

Figure~\ref{cap:lam} shows the tidal parameter $\lambda$
[eq.~(\ref{eq:lamdef})] as a function of $\eta'$.
At $\eta'\lesssim 10$, $\lambda$ is scarcely reduced below its
long-period limit because the $r^8$ factor in eq.~(\ref{eq:lamdef})
emphasizes the outer parts of the convection zone where the local
timescale $\tau\ll t_f$.
The asymptotic scaling $\lambda\propto\eta'^{-2}$ is approached
only gradually; near $\eta'=10^2$, $\lambda\propto\eta'^{-1.5}$.
Another salient feature of this plot is the relative size
of $\lambda$ in fully convective versus solar-type stars.
Fully convective stars enjoy an advantage of almost an order
of magnitude in the long-period limit because they are less
centrally condensed; however, this advantage almost disappears
for $\eta\gg 10$ because the fully convective star has a longer
local timescale $\tau_l(r)$ than the solar-type star at the same
fractional radius.
To identify the point corresponding to a period of 10 days,
we have assumed a fully convective protostellar sun with
$T_{\rm eff}\approx 4070~\mbox{K}$, following \cite{ZB:89},
and find $\eta'\approx 160$.

Fig.~\ref{cap:lam} shows that for solar-type stars, $t_{\rm circ}$ is
increased by $\approx 13(10~\mbox{d}/P)^{1.5}$ with respect
to the estimate (\ref{eq:tcirc0}).
The corresponding $P_{\rm circ}$
of old solar-type stars should be $\approx 1.3~\mbox{d}$ instead of the
value given in eq.~(\ref{eq:Pcirc0}), and still less
for stars whose main-sequence lifetimes are longer than 16~Gyr.

This figure also implies that $\lambda(\eta')\approx\lambda(0)$ for
the giant-star binaries studied
by Verbunt \& Phinney (1996, henceforth VP).
The friction time $t_f\propto T_{\rm eff}^{-4/3}$
at fixed mass [eq.~(\ref{eq:tfric}),
but $T_{\rm eff}$ varies only very slowly along the giant branch: 
VP quote $T_{\rm eff}/T_{\rm eff,\sun}\approx(R/R_{\sun})^{-0.11}$.
Even at the tip of the giant branch ($R\approx 160 R_{\sun}$),
$2\pi t_f\approx 5.7~\mbox{yr}$,
whereas the binary periods of interest are of order
$10^{2.5}-10^{3.3}~\mbox{d}$, so that $\eta'\lesssim 6$.
Thus turbulent viscosity is expected to be fully efficient in these binaries,
and the results of our paper are consistent with the good agreement
VP found between their giant-star binaries and the long-period limit
of Zahn's theory.

\subsection{Pre-main-sequence circularization}

In summary, 
{\it one cannot explain the circularization of old main-sequence binaries
by turbulent convective viscosity.}
This is true even if one ignores the reduced efficiency of turbulent
dissipation at short tidal periods.   If taken into account, as our
results would indicate that it must be,
the inefficiency widens the discrepancy between the
observed and computed circularization period
to a full order of magnitude.

\cite{ZB:89} (henceforth ZB)
have suggested that circularization occurs primarily
{\it before} the main sequence.
Although the T Tauri phase  lasts $\lesssim 10^7~\mbox{yr}$,
it may dominate the circularization process because the stellar radius
is larger during this phase than on the main sequence
(cf \cite{SW:93}).
Other things being equal, the circularization rate varies as $R^8$
[eq.~(\ref{eq:crate0})].
Hence a factor $\approx 3$ in radius may compensate for a factor
$\approx 10^{-4}$ in lifetime.

ZB have integrated the circularization rate against time for low-mass
T Tauri stars following the evolutionary tracks of Stahler and
collaborators (\cite{S:83}, \cite{S:88}).
According to ZB's results, circularization during the
pre-main-sequence (PMS) phase is effective out to periods of 
$7.2-8.5~\mbox{d}$ for stellar masses $0.5-1.25 M_{\sun}$.
This is based on the linear prescription (\ref{eq:Zred})
for the inefficiency factor.
ZB remark that the
time-averaged circularization rate is $2-5$ times smaller than
it would have been if they had evaluated the turbulent
viscosity in the long-period limit.
We believe, however, that the quadratic prescription (\ref{eq:Gred})
should be used instead of the linear one.
For ZB's initial protosun ($M=M_{\sun}$, $R= 4.79 R_{\sun}$, 
$T_{\rm eff}= 4070~\mbox{K}$),
the dashed curve in Fig.~\ref{cap:lam} indicates a reduction factor
of $130$.
Although we have not repeated ZB's evolutionary calculations,
we can rescale their results to obtain a new prediction for
$P_{\rm circ}$ by assuming
that our initial reduction factor ($130$) should replace
their maximum reduction factor ($5$).
Since $\lambda\propto \eta'^{-1.5}$ over the range of present interest,
eq.~(\ref{eq:crate0}) implies $t_{\rm circ}\propto P^{3.83}$, and so
\begin{equation}
P_{\rm circ,PMS}\approx \left(\frac{130}{5}\right)^{-1/3.83}~8~\mbox{d}
\approx 3.4~\mbox{d}
\end{equation}
Given ZB's assumptions about the protosun, therefore, we agree with
their conclusion that circularization by turbulent convection is much
more effective in the T Tauri phase than on the main sequence,
although we have halved their prediction for $P_{\rm circ}$.

One can imagine circumstances in which pre-main-sequence
circularization would be much more effective.
The evolution of protostars is still uncertain; perhaps protostellar
radii are briefly larger than Stahler and ZB suppose.
Tidal interaction with an intrabinary or circumbinary disk may be
important, though in some cases the interaction may {\it increase}
eccentricity (\cite{ACLP}, \cite{AL:94}).
Observations of PMS spectroscopic binaries are difficult to interpret.
Difficulties arise from the paucity of 
known short-period orbits, but perhaps also from genuine variety in the
physical circumstances of PMS systems.
\cite{M:94} concludes that the evidence to date
indicates $P_{\rm circ}\approx 4~{\rm d}$ among
PMS binaries with solar-mass components.
This agrees, perhaps fortuitously, with our revision of ZB's theory.
At any rate, this is much shorter than the conservative inference
$P_{\rm circ}\gtrsim 12~\mbox{d}$ from observations of old halo binaries.

Perhaps the PMS history of the low-metallicity halo binaries was
systematically different from that of present-day T-Tauri binaries
in such a way as to enhance PMS circularization of the former---e.g.
longer-lived or more massive disks.
This could give rise to a spurious correlation between $P_{\rm circ}$
and the age (or metallicity) of the binary population.
Evidence against this idea is that solar-mass binaries in
the relatively young Pleiades and Hyades clusters show intermediate
transition periods ($7.05$ and $8.5~\mbox{d}$, \cite{MDL:92}
and references therein).

Pending more data for short-period binaries of all ages,
it seems that gradual circularization does proceed on the main sequence, but
that some mechanism other than turbulent convection must be responsible.

\acknowledgments
We would like to acknowledge much helpful and patient advice
from Bohdan Paczynski concerning the theoretical evolution
and observed properties of stars in general and of binaries
in particular.  This work was supported by NASA under grant NAG5-2796.





\begin{thebibliography}{}

\bibitem[Argyris, Faust \& Haase 1994]{AFH:94}
Argyris, J., Faust, G., \& Haase, M. 1994, An Exploration of Chaos,
(Amsterdam; New York: North Holland)

\bibitem[Artymowicz et al. 1991]{ACLP}
Artymowicz, P., Clarke, C. J., Lubow, S. H., \& Pringle, J. E. 1991,
\apj, 370, L35

\bibitem[Artymowicz \& Lubow 1994]{AL:94}
Artymowicz, P. \& Lubow, S. H. 1994, \apj 421, 651

\bibitem[Bahcall, Pinsonneault, \& Wasserburg (1995)]{BP:95}
Bahcall, J. N., Pinsonneault, M. H., \& Wasserburg, G. J. 1995,
Rev. Mod. Phys., 67, 781

\bibitem[Bergbusch \& VandenBerg (1992)]{BV:92}
Bergbusch, P. A. \& VandenBerg, D. A. 1992, \apjs, 81, 163

\bibitem[Goldman \& Mazeh (1991)]{GM:91}
Goldman, I. \& Mazeh, T. 1991, \apj, 376, 260

\bibitem[Goldreich \& Keely (1977)]{GK:77}
Goldreich, P. \& Keely, D. A. 1977, \apj, 211, 934

\bibitem[Goldreich \& Kumar (1988)]{GK:88}
Goldreich, P. \& Kumar, P. 1988, \apj, 326, 462

\bibitem[Goldreich, Kumar, \& Murray (1994)]{GKM:94}
Goldreich, P., Kumar, P., \& Murray, N. 1994, \apj, 424, 466


\bibitem[Goldreich \& Nicholson (1977, henceforth GN)]{GN:77}
Goldreich, P. \& Nicholson, P. D. 1989, Icarus, 30, 301 (GN)

\bibitem[Goldreich \& Nicholson (1989)]{GN:89}
Goldreich, P. \& Nicholson, P. D. 1989, \apj, 342, 1079

\bibitem[Gonczi (1982)]{Gon:82}
Gonczi, G. 1982, \aap, 110, 1

\bibitem[Kumar \& Goodman (1996)]{KG}
Kumar, P., and Goodman, J. 1996, \apj, 466, in press

\bibitem[Laird, Carney, \& Latham 1988]{LCL:88}
Laird, J. B., Carney, B. W., \& Latham, D. W. 1988, \aj, 95, 1843

\bibitem[Landau \& Lifshitz 1980]{LL:80}
Landau, L. D. \& Lifshitz, E. M. 1980, Statistical Physics,
Part I, 3rd ed. (Pergamon: Oxford)

\bibitem[Latham et al. (1992a)]{L:92}
Latham, D. W., et al. 1992, \aj, 104, 774

\bibitem[Latham et al. (1992b)]{LMMD}
Latham, D. W., Mathieu, R. D., Milone, A. A. E., \& Davis, R. J. 1992, 
in Binaries as Tracers of Stellar Evolution,
eds. A. Duquennoy \& M. Mayor (Cambridge: Cambridge U Press), p. 132

\bibitem[Mathieu (1994)]{M:94}
Mathieu, R. D. 1994, \araa, 32, 465

\bibitem[Mathieu \& Mazeh 1988]{MM:88}
Mathieu, R. D. \& Mazeh, T. 1988, \apj, 326, 256

\bibitem[Mathieu, Latham, \& Griffin 1990]{MLG:90}
Mathieu, R. D., Latham, D. W., \& Griffin, R. F. 1990, \aj, 100, 1859

\bibitem[Mathieu et al. 1992]{MDL:92}
Mathieu, R. D., Duguennoy, A., Latham, D. W., Mayor, M.,
Mazeh, T., \& Mermilliod, J.-C. 1992,
in Binaries as Tracers of Stellar Evolution,
eds. A. Duquennoy \& M. Mayor (Cambridge: Cambridge U Press), p. 278

\bibitem[Nelson \& Tremaine 1997]{NT:97}
Nelson, R. W., and Tremaine, S. 1997, in preparation

\bibitem[Pedlosky 1987]{Pedlosky}
Pedlosky, J. 1987, Geophysical Fluid Dynamics, 2nd Edition,
(New York: Springer)

\bibitem[Phinney 1992]{P:92}
Phinney, E. S., 1992, Phil. Trans. R. Soc. London, Ser. A., 341, 39

\bibitem[Stahler 1983]{S:83}
Stahler, S. W. 1983, \apj, 274, 822

\bibitem[Stahler 1988]{S:88}
--- 1983, \apj, 332, 804

\bibitem[ Stahler \& Walter 1993]{SW:93}
Stahler, S. W. \& Walter, F. M. 1993, in Protostars and Planets III,
eds. E. H. Levy \& J. I. Lunine, p. 405

\bibitem[Spruit (1996)]{Spruit:96}
Spruit, H. C. 1996, Mem. Soc. Astron. It., to appear; SISSA e-print
9605020

\bibitem[Verbunt \& Phinney 1996]{VP:96}
Verbunt, F., \& Phinney, E. S. 1996, \aap, 296, 709 (VP)

\bibitem[Zahn 1966]{Z:66}
Zahn, J.-P. 1966, Annals d'Astrophysique, 29, 489

\bibitem[Zahn 1970]{Z:70}
Zahn, J.-P. 1970, \aap, 4, 452

\bibitem[Zahn 1975]{Z:75}
Zahn, J.-P. 1975, \aap, 41, 329

\bibitem[Zahn 1977]{Z:77}
Zahn, J.-P. 1977, \aap, 57, 383

\bibitem[Zahn 1989]{Z:89}
Zahn, J.-P. 1989, \aap, 220, 112 (Z89)

\bibitem[Zahn 1992]{Z:92}
Zahn, J.-P. 1992, in Binaries as Tracers of Stellar Evolution,
eds. A. Duquennoy \& M. Mayor (Cambridge: Cambridge U Press), p. 253

\bibitem[Zahn \& Bouchet (1989)]{ZB:89}
Zahn, J.-P. \& Bouchet, L. 1989, \aap, 223, 112 (ZB)

\end{thebibliography}
\end{document}